\documentclass[12pt]{iopart}

\usepackage{iopams}
\usepackage{amsfonts,amssymb}
\usepackage{color,calc}
\usepackage[dvips]{graphicx}
\usepackage{bm}
\usepackage{citesort}

\def\be{\begin{equation}}
\def\ee{\end{equation}}
\def\bea{\begin{eqnarray}}
\def\eea{\end{eqnarray}}
\def\bse{\begin{subequations}}
\def\ese{\end{subequations}}

\def\ket#1{\vert #1 \rangle}
\def\bra#1{\langle #1 \vert}

\def\HRed{\hat{H}_{\mathrm{I,R}}}
\def\HBlue{\hat{H}_{\mathrm{I,B}}}

\def\phonons{\nu}

\def\totalarea{A_\mathrm{tot}}
\def\wtrap{\omega_\mathrm{tr}}
\def\noon{\mathrm{N}}
\def\tot{\mathrm{tot}}

\begin{document}

\title[Creation of arbitrary Dicke and NOON states by global addressing]{Creation of arbitrary Dicke and NOON states of trapped-ion qubits by global addressing with composite pulses}
\author{Svetoslav S. Ivanov$^{1,2}$, Nikolay V. Vitanov$^{2}$ and Natalia V. Korolkova$^{1}$}
\address{$^1$ School of Physics and Astronomy, University of St. Andrews, North Haugh, St. Andrews, Fife, KY16 9SS, Scotland}
\address{$^2$ Department of Physics, Sofia University, James Bourchier 5 Blvd, 1164 Sofia, Bulgaria}
\eads{\mailto{ssi@st-andrews.ac.uk},
\mailto{sivanov@phys.uni-sofia.bg}}

\date{\today}

\begin{abstract}
We propose a fast and efficient technique to create classes of highly entangled states of trapped ions, such as arbitrary Dicke states and superpositions of them, e.g. NOON states.
The ions are initialized in the phonon ground state and are addressed globally with a composite pulse that is resonant with the first motional sideband.
The technique operates on comparatively short time scales, as resonant interactions allow one to use the minimum laser pulse area. The number of single pulses from the composite sequence is equal to the number of ions, thus the implementation complexity grows only linearly with the size of the system. The approach does not require individual addressing of the ions in the trap and can be applied both inside and outside the Lamb-Dicke regime.
\end{abstract}


\maketitle

\section{Introduction}

Entanglement is the most distinctive feature of quantum states involving many particles. Within the framework of quantum information science,
it may be viewed as a resource for the processing of information in ways not permitted by classical logic \cite{Chuang}.
Entanglement has various physical applications such as dense coding, quantum teleportation, quantum cryptography, quantum metrology, etc. \cite{entanglement}, which are essential for quantum communication and information processing (QIP). It is indisputable that entanglement plays a key role in QIP as quantum computers are implemented by many-body systems, generally characterized by multi-partite entangled states. Since the primary resource for quantum computation is a Hilbert-space dimension, which grows exponentially with the available physical resources \cite{Kohout}, the benefits of a quantum over a classical computation increase with the size of the physical system. This has inspired an intensive research aiming to create and study the properties of multi-partite entangled states.

A very prominent class of such states are Dicke states $\ket{W^N_n}$, originally introduced in \cite{Dicke}. They contain a given number of excitations $n$ (qubits in state $\ket{1}$) shared evenly amongst all $N$ qubits:
\be
\ket{W^N_n} = \frac{1}{\sqrt{C^N_n}}
\sum_{k}P_k \ket{\underbrace{1,\ldots,1}_n,\underbrace{0,\ldots,0}_{N-n}},
\ee
where $P_k$ denote the set of all permutations of the excitations and $C^N_n\equiv N!/\left[n!(N-n)!\right]$. Notably, Dicke states are immune against collective dephasing, which is a dominant source of decoherence in various systems, such as trapped ions \cite{DFS}. Therefore, while still offering exponential dimensionality (of $C^N_{N/2}\approx 2^N/\sqrt{\pi N/2}$, example for $n=N/2$), the Dicke manifold can be used as a decoherence-free computational subspace, as in \cite{Grover}. Dicke states generalize $W$ states, which can be used for quantum communication \cite{Communication}. Furthermore, Dicke states exhibit genuine multi-partite entanglement \cite{DickeEntanglement}, which is robust against particle loss and is highly resilient vs external perturbations and measurements on single qubits \cite{Robust}. Thus, Dicke states can serve as a versatile resource for the preparation of multiparticle entangled states; through projective measurements on some of the qubits one can obtain states from various entanglement classes. Due to their robust entanglement, these states are particularly well suited
for the experimental examination of multi-partite entanglement and can be used to test fundamental concepts of quantum mechanics.

Theoretical proposals exist for the generation of Dicke states in a number of physical systems, including ensembles
of neutral atoms \cite{prop1}, trapped ions \cite{Ian,prop2}, quantum dots \cite{prop3} and using linear optics \cite{prop4}. Of these ion traps are perhaps the best suited for their unparalleled level of experimental control. We notice, however, that the existing trapped-ion proposals possess one or a combination of the following drawbacks: (1) they cannot create arbitrary but only particular Dicke states, and thus do not offer a general approach; (2) individual ion addressing is required, which poses significant experimental challenges to scalability; (3) the number of physical interactions needed scales very fast with the system size; (4) initialization in a particular Fock state is required; (5) Dicke states are achieved with some probability and post selection is required; (6) adiabatic techniques require in general very long interaction times. Consequently, extending the proposed techniques for larger system sizes remains a formidable challenge.

In this paper, we take a different approach and propose a simple, general and very efficient technique for the creation of a large class of highly entangled states in systems of trapped ions. These can be arbitrary Dicke states and any superposition of these, such as NOON states, which are invariant, up to a phase, under the exchange of any two ions. Besides its generality, the proposed technique is particularly advantageous due to several features.
It uses composite pulse sequences -- a series of laser pulses, each with a particular area and phase.
Composite pulses are a conceptually simple and very powerful control tool, which enjoys large popularity in experimental physics. Though they were first developed for the needs of NMR, they were successfully applied in trapped-ion systems where many major accomplishments have been made, for example in the field of quantum information processing \cite{cpulses}. Their simplicity and efficacy in controlling quantum systems stem from the basic physical notion of interference. Thus, using specially designed composite pulses, our technique requires much fewer interaction steps compared to the traditional approaches, exploiting quantum circuits of a vast number of concatenated one- and two-qubit gates: the number of pulses in our approach is equal to the number of ions. Therefore it grows only linearly with the system size, thus offering only moderate levels of experimental complexity. Another advantage of our method is that it assumes collective interaction with all ions and does not require to manipulate exclusively individual ions or pairs of such with focused laser beams, which often presents a principal experimental challenge. The laser fields are resonant with the first motional sideband transition of the ions, which results in short interaction times, as opposed to adiabatic techniques. Our technique is applicable also outside the Lamb-Dicke regime. This offers the potential to overcome various detrimental effects, such as light shifts and off-resonant excitations \cite{Steane2000}, which might occur in experimental implementations.

\section{Model}

\subsection{Hamiltonian}
We consider $N$ ions confined in a linear Paul trap, which are cooled to their vibrational ground state.
Each ion has two relevant internal states $\ket{0}$ and $\ket{1}$, with respective transition frequency $\omega_{0}$.
The linear ion crystal interacts uniformly with a laser pulse tuned on one of the sidebands of the (longitudinal) center-of-mass mode, with
frequency $\omega_{\mathrm{L}}=\omega_{0}\pm\wtrap$, where $\wtrap$ is the axial trap frequency.
The plus sign stands for the blue sideband, while the minus sign is for the red sideband.
After making the optical and vibrational rotating-wave approximations, the interaction Hamiltonian in the interaction representation
for the red and blue sidebands, respectively, has the form \cite{Leibfried2003}
\numparts
\begin{eqnarray}
\label{HamiltonianR}
\HRed &=& \frac{1}{2}\hbar g(t) \hat{a}(\eta)\hat{J}_{+} + H.c.,\\
\label{HamiltonianB}
\HBlue &=& \frac{1}{2}\hbar g(t) \hat{a}^\dagger(\eta)\hat{J}_{+} + H.c.,
\end{eqnarray}
\endnumparts
where $g(t)=\eta\Omega(t)\exp(-\eta^2/2)/\sqrt{N}$ is the coupling of the internal atomic states to the vibrational mode, producing pulse area $A=\int_{-\infty}^{\infty}g(t)\rmd t$, $\eta$ is the single-ion Lamb-Dicke parameter, $\Omega(t)$ is the real-valued time-dependent Rabi frequency.
Here $\hat{a}(\eta)=\sum_{n=0}^\infty (n+1)^{-1/2} L^1_n(\eta^2)\ket{n}\bra{n+1}$ and $\hat{a}^\dagger(\eta)=\left[\hat{a}(\eta)\right]^\dagger$ are the phonon lowering and raising operators, $L^a_n(x)$ being the generalized Laguerre polynomial. $\hat{J}^{\pm}=\sum_{k=1}^{N} e^{\pm i\varphi_k}\sigma_k^{\pm}$,
where $\varphi_k$ is the phase of the laser field interacting with the $k$th ion, and $\sigma^{+}_k=\ket{1_k}\bra{0_k}$ and $\sigma^{-}_k=\ket{0_k}\bra{1_k}$ are the raising and lowering operators for the internal states of the $k$th ion.
In the Lamb-Dicke limit the operators $\hat{a}^\dagger(\eta)$ and $\hat{a}(\eta)$ become the ordinary creation and annihilation operators of the center-of-mass phonons.

We now perform the transformation $\ket{1_k}\rightarrow\ket{\widetilde{1}_k}e^{-i\varphi_k}$, thereby incorporating the laser phase into the atomic states. As a result, the Hamiltonian is recast in terms of the usual pseudospin operators $\hat{J}^{\pm}=\sum_{k=1}^{N} \widetilde{\sigma}_k^{\pm}$.
The energy pattern splits into manifolds corresponding to $n$ atomic and $\phonons$ motional excitations, Fig. \ref{fig1}(a).
For red-sideband interaction we have $n+\phonons=m_\mathrm{R}$, i.e. the total number of quanta $m_\mathrm{R}$ is conserved, while for blue sideband we have $n-\phonons=m_\mathrm{B}$, i.e. the difference of the quanta is conserved.
To create symmetric entangled states a suitable choice is $m_\mathrm{R}=N$ and $m_\mathrm{B}=0$. Though our method is equally applicable for both, in what follows we will assume blue-sideband interaction.

Figure \ref{fig1}(a) depicts all states, which are accessible if one starts from the ground state $\ket{000}\ket{0}$. The example given is for a chain of $N=3$ ions. The ions interact with a blue-sideband laser field, $\omega_\mathrm{L}=\omega_0+\wtrap$, which couples equally each ion's internal state to the collective motional center-of-mass mode: $\ket{0_k}\ket{\phonons}\leftrightarrow\ket{1_k}\ket{\phonons+1}$. The system is described by the anti-Jaynes-Cummings Hamiltonian (\ref{HamiltonianB}). To this end, we adopt the wavefunction notation $\ket{\psi} \ket{\phonons}$, where $\ket{\psi}=\ket{q_1q_2\cdots q_{n+1}}$ is the collective internal state of the ion qubits, with $q_k=0$ or 1, and $\ket{\phonons}$ is the vibrational Fock state of $\phonons$ phonons.

\subsection{Hilbert space factorization}

\begin{figure}[tb]
\begin{center}
\includegraphics[angle=0,width=0.55\columnwidth]{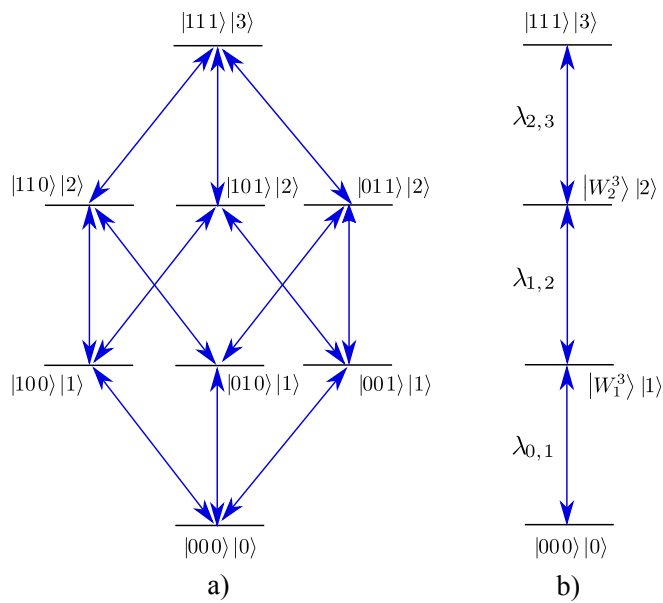}
\end{center}
\caption{
(a) Linkage pattern of the collective states of $N=3$ trapped ions equally coupled to their common center-of-mass mode
by a uniform laser field.
The difference between the number of ionic excitations $n$ and the number of vibrational phonons $\phonons=n$ is conserved.
The laser beam is tuned to the
blue-sideband resonance, $\omega_\mathrm{L}=\omega_0+\wtrap$.
(b) Chains comprising symmetric Dicke states $\ket{W^N_n}\ket{\phonons}$ for $N=3$. These constitute the whole set of states that can be accessed by a uniform laser driving starting from the state $\ket{00\cdots 0}\ket{0}$. The states are coupled resonantly on the first blue sideband with $\lambda_{\phonons-1,\phonons}(t)$ being the coupling strengths.}
\label{fig1}
\end{figure}

In order to study the dynamics of our system, it is convenient to introduce a new basis, which consists of the set of eigenvectors of the two commuting
pseudospin operators $\hat{J}^2$ and $\hat{J}_{z}$, where $\hat{J}^2=\frac{1}{2}(\hat{J}_{+}\hat{J}_{-}+\hat{J}_{-}\hat{J}_{+})+\hat{J}_z^2$.
Each state is assigned two quantum numbers, $j$ and $m_j$, respectively. Since $\hat{J}^2$ commutes with the Hamiltonian,
the Hilbert space factorizes into a set of decoupled chains with different values of $j$; the Hamiltonian preserves $j$.
The meaning of this becomes more transparent if we notice that $\hat{J}^2=\sum_{k,l=1}^N \frac{1}{2} S_{kl}-\frac{1}{4}(-1)^{\delta_{kl}}\mathbf{1}$, with
$S_{kl}=\sigma^{+}_k\sigma^{-}_l+\sigma^{-}_k\sigma^{+}_l+\frac{1}{2}\sigma_{k,z}\sigma_{l,z}+\frac{1}{2}(-1)^{\delta_{kl}}\mathbf{1}$ denoting the action of the swapping operator,
which exchanges ions $k$ and $l$. Therefore, each $j$ stands for a particular symmetry with respect to exchanging ions. Since our class of target
states is invariant under the action of each $S_{kl}$, it comprises the eigenstates of $S_{kl}$ with unit eigenvalue. Thus the eigenvalue of $\hat{J}^2$ is $(1+N/2)N/2$. Hence, the chain containing
our states is assigned $j=N/2$, and by analogy with the traditional angular momentum operators, the number of states is equal to $2j+1=N+1$.
These are all Dicke states $\ket{W^N_n}\ket{\phonons}$, with $n$ being the number of atomic excitations: $n=0,\ldots,N$. Hence, all states we
are interested in are contained in a single chain, and are coupled by the Hamiltonian (\ref{HamiltonianB}) in the order given (Fig. \ref{fig1}(b)). As long as the interaction with the ions is distributed uniformly, the dynamics is enclosed in this chain as it gets decoupled from the rest of the Hilbert space.
The states differ by energy, which is measured by the operator $\hat{J}_{z}$. Its eigenvalues $m_j$ vary from $-j$ to $j$ and define the number of excited ions, $n=j+m_j=N/2+m_j$.

For the following analysis, it will be necessary to
go further and calculate the coupling coefficients in the new basis.
The coupling between the
neighbors $\ket{j,m_j}$ and $\ket{j,m_j-1}$ follows immediately
from the matrix elements of the operators $\hat{a}^{\dag}(\eta)$, $\hat{a}(\eta)$ and
$\hat{J}_{\pm}$:
\be
\lambda_{\phonons-1,\phonons}(t) = g(t) L^1_{\phonons-1}(\eta^2)\sqrt{N-\phonons+1}.
\label{coupling}
\ee

In the following, for conciseness we will consider only operation inside the Lamb-Dicke regime, which requires $\eta\ll 1$.

\section{Implementation}

Our method begins with the initialization of the string of $N$ ions in the collective internal and vibrational ground state $\ket{\psi}\ket{\phonons}=\ket{00\cdots 0}\ket{0}$ \cite{comment1}. For ease of notation, the indices in $\ket{\psi}$ denoting states will be omitted hereafter.

\subsection{Creation of Dicke states}

The $N$-ion Dicke states $\ket{W^N_n}\ket{\phonons}$ are constructed in the following way.
A series of $N$ pulses is applied globally on all ions, each pulse having a particular area $A_k$ and phase $\phi_k$. The pulses are resonant with the first blue sideband relative to the center-of-mass (COM) mode, i.e. the carrier frequencies are $\omega_\mathrm{L}=\omega_0+\wtrap$. Thereby the state $\ket{00\cdots 0}\ket{0}$ is coupled to all states $\ket{W^N_n}\ket{n}$, shown in Fig. \ref{fig1} (b). If other modes are used, one would connect states of another symmetry. Because resonant interactions are employed, the dynamics is defined only by the pulse areas $A_k$ and does not depend on the temporal pulse shape.

To study the effect of this interaction we derive the propagator $U(A_k,\phi_k)$ describing the dynamics of the chain of Dicke states (Fig. \ref{fig1}(b)), subject to a laser pulse of area $A_k$ and phase $\phi_k$. This is done by exact diagonalization and exponentiation of the Hamiltonian, $U(A_k,\phi_k)=\mathrm{exp}\left(-\frac{i}{\hbar} \int \HBlue \rmd t\right)$, with the coupling in $\HBlue$ being phased, $g(t)\rightarrow g(t)e^{i\phi_k}$. The total sequence of $N$ pulses, having area $\totalarea=\sum_{k=1}^N A_k$, is represented by the propagator
\be
U_\tot=U(A_N,\phi_N)U(A_{N-1},\phi_{N-1})\cdots U(A_1,\phi_1).
\label{propagator}
\ee
We fix $\phi_1=0$, which defines our phase reference. Hence, the total propagator $U_\tot$ is defined by $N$ pulse areas and $N-1$ phases, a total of $2N-1$ variables, which can be varied as free parameters.

Dicke states, or various superpositions of these, are obtained for specific sets of parameters, which are determined numerically through maximizing the fidelity with the target state $\ket{t}$, seeking unity. The fidelity is defined as
\be
\mathcal{F}(A_k,\phi_k,t)=\left|\bra{t}U_\tot \ket{00\cdots 0}\ket{0}\right|^2
\ee
and is a function of all areas $A_k$ and phases $\phi_k$. The numerical optimization procedure runs over the $2N-1$ dimensional space of $A_k$ and $\phi_k$ and follows Newton's gradient-based method. Because this is a local optimization algorithm, we iteratively pick the initial parameter values using the Monte-Carlo scheme. Out of the many solutions obtained, we select the one having the minimal total pulse area $A_\tot$.

The numerical optimization is computationally not difficult even beyond $N=15$ ions, even though the dimension of the Hilbert space scales
exponentially with $N$. The reason is that the system resides only in the chain of symmetric states of dimension $N+1$, shown in Fig. \ref{fig1} (b).

\begin{figure}[tb]
\begin{center}
\includegraphics[angle=0,width=0.60\columnwidth]{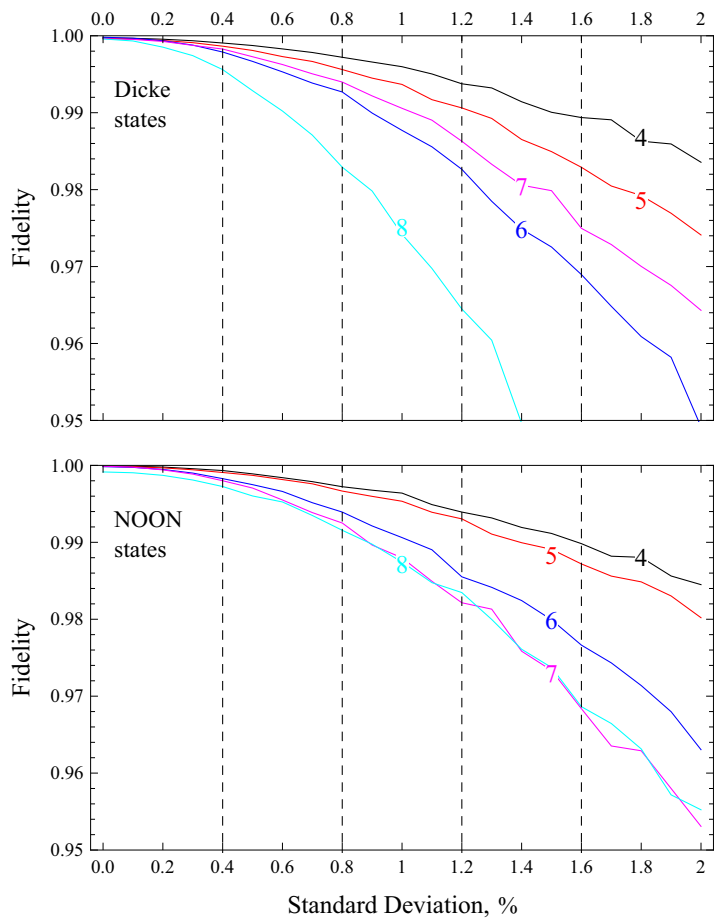}
\end{center}
\caption{
Fidelity for the creation of various Dicke (top) and NOON states (bottom) vs the standard deviation in the control parameters $A_k$ and $\phi_k$, with $k=1,\ldots, N$. The parameters are listed in Tables \ref{table1} and \ref{table2}, respectively, and the numbers denote the number of ions $N$.}
\label{error}
\end{figure}

In Table \ref{table1} we provide examples of pulse sequences, which yield Dicke states $\ket{W^N_n}$ for different number of ions $N$.
We choose $n=\lfloor N/2\rfloor$, $\lfloor x\rfloor$ being the integer part of $x$.
In a real experiment one may not be able to set the control parameters exactly as prescribed by Table \ref{table1}. The fluctuations around the optimal values would result in a decrease of the fidelity. We have investigated this scenario and the result is shown in Figure \ref{error} (top), which illustrates the final fidelity for different Dicke states vs the standard deviations of the control parameters $A_k$ and $\phi_k$. It is noteworthy that the calculated fidelity stays well above 95\% for deviations of the order of 1\%, which are typical for the present state-of-the-art technology \cite{Haffner}.

\begin{table}[tb]
\begin{center}
\begin{tabular}{|l l l|}
\hline\multicolumn{3}{|c|}{Dicke states, $\ket{W^N_n}$} \\ \hline
\hline
$N$ & $A_\tot$ & ($A_1$, $\phi_1$; $A_2$, $\phi_2$; \ldots; $A_N$, $\phi_N$) \\ \hline
3    &2.53& (0.369, 0; 0.484, 2.39; 1.682, 2.976)\\
4    &2.28& (0.805, 0; 0.495, 1.728; 0.793, 0.566; 0.191, 0.079) \\
5    &2.11& (0.795, 0; 0.278, 0.403; 0.480, 0.075; 0.223, 0.309; \\
     &    &  0.333, 0.915) \\
6    &2.12& (0.562, 0; 0.315, 1.478; 0.343, 0.854; 0.277, 0.417; \\
     &    &  0.126, 0.091; 0.501, 1.423) \\
7    &2.15& (0.107, 0; 0.584, 1.694, 0.562, 1.566, 0.497, 1.313, \\
     &    &  0.039, 1.956, 0.158, 1.301, 0.206, 1.847) \\
8    &2.46& (0.539, 0; 0.216, 0.389; 0.459, 0.098; 0.251, 1.560; \\
     &    &  0.464, 0.816; 0.25, 0.388; 0.25, 2.078; 0.03, 1.607) \\
9    &3.35& (0.51, 0; 0.234, 0.83; 0.9, 0.304; 0.19, 2.025; 0.352, \\
     &    &  0.164; 0.379, 0.556; 0.358, 0.097; 0.199, 0.239; \\
     &&      0.231, 0.471) \\
10   &3.89& (0.621, 0; 0.367, 1.147; 0.097, 0.994; 0.616, 1.709; \\
     &&      0.113, 0.263; 0.203, 0.661; 0.579, 0.328; 0.223, 0.831; \\
     &&      0.775, 0.909; 0.292, 0.462) \\
\hline
\end{tabular}
\caption{Exemplary areas $A_k$ and phases $\phi_k$ (in units of $\pi$) for composite pulse sequences,
which produce $N$-ion Dicke states $\ket{W^N_n}$.
We choose $n=\lfloor N/2\rfloor$,
where $\lfloor x\rfloor$ is the integer part of $x$.  The composite sequences are described by the propagator
(\ref{propagator}) and comprise $N$ phased pulses, tuned on the first blue sideband. It is noteworthy that
for increasing values of $N$ the total pulse area $A_\tot$ is less than $(N/2)\pi$ (we have checked that this
property holds also for $N>10$).
}
\label{table1}
\end{center}
\end{table}

\subsection{Creation of NOON states}

We can also create arbitrary superpositions of the states contained in the chain of accessible states, shown in Fig. \ref{fig1}(b).
Of particular interest is the possibility to generate NOON states $\ket{\noon^N}$, which are another very important class of highly nonclassical entangled
states. They can be defined as an equal-probability superposition of two Dicke states $\ket{W^N_n}$, whereby the excitation is contained either in the
internal state or in the motional state of the ions:
$\ket{\noon^N}=\ket{W^N_N}\ket{0}+\ket{W^N_0}\ket{N}\equiv \ket{11\cdots 1}\ket{0}+\ket{00\cdots 0}\ket{N}$ (for simplicity, normalization constants are
omitted throughout). The ions are maximally correlated as measuring the state of one ion determines the state of all $N$ ions.

Various applications of NOON states of both fundamental and practical interest have been suggested, such as entanglement enhanced metrology and sub-wavelength lithography \cite{Bollinger,NOON}. Different schemes for the creation of NOON states in trapped-ion systems have been proposed theoretically \cite{Bollinger,NOONth} and realized experimentally \cite{NOONexp}. However, they are subject to requirements for individual ion addressing, limited number of ions or long interaction times where adiabaticity is employed.

Unlike the previous proposals, our method allows us, by using global addressing, to create arbitrary NOON states. As the interaction is resonant,
the states are created on short time scales. We perform the same manipulation as for the creation of Dicke states --
a sequence of laser pulses is applied, each addressing globally the chain of ions and having a particular
area $A_k$ and a relative phase $\phi_k$. The pulses are
resonant with the first blue-sideband transition. For particular pulse sequences one obtains the coherent superposition
$\ket{W^N_0}\ket{0}+\ket{W^N_N}\ket{N}$. If we logically interchange $\ket{0}$ and $\ket{1}$, $\ket{0}\leftrightarrow\ket{1}$,
this state corresponds to the NOON state $\ket{\noon^N}$.

In Table \ref{table2} we provide examples of pulse sequences, which yield NOON states $\ket{\noon^N}$ for different number of ions $N$.
Figure \ref{error} (bottom) illustrates the final fidelity of various NOON states vs the standard deviations of the control parameters $A_k$ and $\phi_k$. As for the Dicke states, the calculated fidelity stays well above 95\% for deviations of the order of 1\%.

\begin{table}[tb]
\begin{center}
\begin{tabular}{|l l l|}
\hline\multicolumn{3}{|c|}{NOON states, $\ket{\noon^N}$} \\ \hline
\hline
$N$ & $A_\tot$ & ($A_1$, $\phi_1$; $A_2$, $\phi_2$; \ldots; $A_N$, $\phi_N$) \\ \hline
3    &1.60& (0.696, 0; 0.640, 1.511; 0.259, 1.962) \\
4    &1.63& (0.402, 0; 0.291, 0.151; 0.667, 1.819; 0.271, 1.465) \\
5    &1.88& (0.494, 0; 0.249, 0.652; 0.651, 1.271; 0.313, 0.806; \\
     &&      0.175,1.175) \\
6    &1.83& (0.284, 0; 0.235, 0.219; 0.099, 0.701; 0.673, 1.178; \\
     &&      0.403, 0.665; 0.136, 1.022) \\
7    &2.06& (0.278, 0; 0.300, 0.266; 0.338, 0.034; 0.541, 1.895; \\
     &&      0.277, 2.138; 0.137, 0.662; 0.187, 0.070) \\
8    &2.33& (0.259, 0; 0.923, 0.209; 0.346, 0.408; 0.428, 1.572; \\
     &&      0.003, 1.705; 0.204, 1.216; 0.003, 2.11; 0.162, 1.543) \\
9    &2.46& (0.395, 0; 0.146, 2.556; 0.186, 1.336; 0.237, 1.854; \\
     &&      0.680, 0.740; 0.452, 1.660; 0.169, 0.862; 0.007,\\
     &&      0.222; 0.186, 1.555) \\
10   &2.93& (0.476, 0; 0.239, 1.247; 0.289, 1.380; 0.256, 0.305; \\
     &&      0.228, 2.021; 0.415, 0.220; 0.388, 0.749; 0.059, 1.718; \\
     &&      0.529, 1.823; 0.047, 0.861) \\
\hline
\end{tabular}
\end{center}
\caption{Exemplary areas $A_k$ and phases $\phi_k$ (in units of $\pi$) for composite pulse sequences,
which produce $N$-ion NOON states $\ket{\noon^N}$.
The sequences are described by the propagator (\ref{propagator}) and comprise $N$ phased pulses, tuned on the first blue sideband.
It is noteworthy that
for increasing values of $N$ the total pulse area $A_\tot$ is below $(N/3)\pi$ (we have checked that this
property holds also for $N>10$).}
\label{table2}
\end{table}

\subsection{Rate of creation of the target states}
As already mentioned, the proposed technique operates on comparatively short time scales.
An estimate for the duration of the composite pulse sequences needed to create our symmetric entangled states is given by
$T_\tot=A_\tot/g$. In order to suppress the excitation of the extraneous phonon modes (other than the center-of-mass mode), we limit the coupling strength from above to $g=\wtrap/10$ \cite{James1998}. If we assume a typical trap frequency of $\wtrap = 4$ MHz, as in \cite{Ferdinand2003},
and $A_\tot \approx 2\pi$ as obtained in the above examples for the creation of $\ket{W^6_3}$ and $\ket{\noon^6}$, we obtain $T_\tot\approx 15\mu$s.
For comparison, in \cite{ian} the Dicke state $\ket{W^6_2}$ is created adiabatically in 400 $\mu$s.

Importantly, as can be seen from Tables \ref{table1} and \ref{table2}, for increasing number of ions $N$ the total pulse area $A_\tot$ stays below $(N/2)\pi$ for Dicke states and below $(N/3)\pi$ for NOON states. Therefore, the duration $T_\tot$ increases only linearly with $N$ and is asymptotically limited by $(N/2)T_\pi$ and $(N/3)T_\pi$, respectively, $T_\pi\approx 8\mu$s being the duration of a $\pi$ pulse.

\section{Conclusions}

We have proposed a simple and efficient technique for the creation of arbitrary collective states of trapped ions, which are symmetric under exchange of any two ions. These can be Dicke states and superpositions of these, such as NOON states. The method uses dedicated composite sequences of phased resonant pulses tuned on the first red or blue motional sideband of the center-of-mass mode. The composite sequences comprise $N$ pulses, $N$ being the number of ions, thus the implementation complexity and duration grow only linearly. This is in contrast to other proposals, which require a rapidly increasing number of elementary gates, demanding exclusive interaction with single ions or pairs of ions. As opposed to previous proposals, the ions are addressed globally, thus individual ion access is unnecessary. Due to the resonant type of interaction and because the required by our method total pulse area is as low as $(N/2)\pi$, the states are created on a comparatively short time scale. The method is applicable also outside the Lamb-Dicke regime.

\ack

The research leading to these results has received funding from the
European Community's Seventh Framework Programme (FP7/2007-2013) under
grant agreement n$^\circ$ 270843 (iQIT) and the Bulgarian NSF grants D002-90/08 and DMU-03/103.


\section*{References}
\begin{thebibliography}{99}

\bibitem{Chuang}
M. A. Nielsen and I. L. Chuang, Quantum Computation and Quantum Information (Cambridge University Press, Cambridge,
England, 2000).

\bibitem{entanglement}
C. H. Bennett and S. J. Wiesner, Phys. Rev. Lett. \textbf{69}, 2881-2884 (1992); K. Mattle, H. Weinfurter, P. G. Kwiat, and A. Zeilinger, Phys. Rev. Lett. \textbf{76}, 4656 (1996); C. H. Bennett, G. Brassard, C. Crepeau, R. Jozsa, A. Peres, and W. K. Wootters, Phys. Rev. Lett. \textbf{70}, 1895 (1993); D. Boschi, S. Branca, F. De Martini, L. Hardy, and S. Popescu, Phys. Rev. Lett. \textbf{80}, 1121 (1998); C. H. Bennett and G. Brassard, Proceedings of the IEEE International Conference on Computers, Systems, and Signal Processing, Bangalore, p. 175 (1984).

\bibitem{Kohout}
R. Blume-Kohout, C. M. Caves, and I. H. Deutsch, Found. Phys. \textbf{32} (2002).

\bibitem{Dicke}
R. H. Dicke, Phys. Rev. \textbf{93}, 99 (1954).

\bibitem{DFS}
D. A. Lidar and K. B. Whaley, Lect. Notes Phys. \textbf{622}, 83 (2003).

\bibitem{Grover}
S. S. Ivanov, P. A. Ivanov, I. E. Linington, and N. V. Vitanov, Phys. Rev. A \textbf{81}, 042328 (2010); S. S. Ivanov, P. A. Ivanov, and N. V. Vitanov, A \textbf{78}, 030301(R) (2008); I. E. Linington, P. A. Ivanov, and N. V. Vitanov, Phys. Rev. A \textbf{79}, 012322 (2009).

\bibitem{Communication}
J. Joo, Y.-J. Park, J. Lee, J. Jang, and I. Kim, J. Korean Phys. Soc. \textbf{46}, 763–-768 (2005);
J. Joo, J. Lee, J. Jang, and Y.-J. Park, arXiv:quant-ph/0204003v2;
H. Buhrman, W. van Dam, P. H{\o}yer, and Tapp, A., Phys. Rev. A \textbf{60}, 2737–-2741 (1999).

\bibitem{DickeEntanglement}
G. T\'oth, J. Opt. Soc. Am. B \textbf{24}, 275 (2007);
A. R. Usha Devi, R. Prabhu, and A. K. Rajagopal, Phys. Rev. Lett. \textbf{98}, 060501 (2007).

\bibitem{Robust}
J. K. Stockton, J. M. Geremia, A. C. Doherty, and H. Mabuchi, Phys. Rev. A \textbf{67}, 022112 (2003);
M. Bourennane, M. Eibl, S. Gaertner, N. Kiesel, C. Kurtsiefer, and H. Weinfurter, Phys. Rev. Lett. \textbf{96}, 100502 (2006);
W. D\"ur, Phys. Rev. A \textbf{63}, 020303(R) (2001).

\bibitem{prop1}
J. K. Stockton, R. van Handel, and H. Mabuchi, Phys. Rev. A \textbf{70}, 022106 (2004);
A. Mandilara, V. M. Akulin, M. Kolar, and G. Kurizki, Phys. Rev. A \textbf{75}, 022327 (2007);
C. Thiel, J. von Zanthier, T. Bastin, E. Solano, and G. S. Agarwal, Phys. Rev. Lett. \textbf{99}, 193602 (2007).

\bibitem{Ian}
I. E. Linington and N. V. Vitanov, Phys. Rev. A \textbf{77}, 010302(R) (2008).

\bibitem{prop2}
A. Retzker, E. Solano, and B. Reznik, Phys. Rev. A \textbf{75}, 022312 (2007);
C. E. L\'{o}pez, J. C. Retamal, and E. Solano, Phys. Rev. A \textbf{76}, 033413 (2007);
H. H\"affner, et al., Nature (London) \textbf{438}, 643 (2005); 
D. B. Hume, C. W. Chou, T. Rosenband, and D. J. Wineland, Phys. Rev. A \textbf{80}, 052302 (2009).

\bibitem{prop3}
X. B. Zou, K. Pahlke, and W. Mathis, Phys. Rev. A \textbf{68}, 034306 (2003).

\bibitem{prop4}
N. Kiesel, C. Schmid, G. Tóth, E. Solano, and H. Weinfurter, Phys. Rev. Lett. 98, 063604 (2007);
C. Thiel, J. von Zanthier, T. Bastin, E. Solano, and G. S. Agarwal, Phys. Rev. Lett. 99, 193602 (2007).

\bibitem{cpulses}
T. Monz et al., Phys. Rev. Lett. \textbf{102}, 040501 (2009);
F. Schmidt-Kaler et al., Nature (London) \textbf{422}, 408 (2003);
S. Gulde et al., Nature \textbf{421}, 48 (2003);
F. Schmidt-Kaler et al., Appl. Phys. B \textbf{77}, 789 (2003);
A. M. Childs and I. L. Chuang, Phys. Rev. A \textbf{63}, 012306 (2000).

\bibitem{Steane2000}
A. Steane, Ch. F. Roos, D. Stevens, A. Mundt, D. Leibfried, F. Schmidt-Kaler, and R. Blatt, Phys. Rev. A \textbf{62}, 042305 (2000).

\bibitem{Leibfried2003}
D. Leibfried, R. Blatt, C. Monroe, and D. Wineland, Rev. Mod. Phys \textbf{75}, 281 (2003).

\bibitem{comment1}
This does not necessarily impose the LD regime, which is accessed when both the phonon number and the LD parameter are very small.

\bibitem{Haffner}
H. H\"affner, C.F. Roos, R. Blatt, Phys. Rep. \textbf{469}, 155-203 (2008).


\bibitem{Bollinger}
J. J. Bollinger, W. M. Itano, D. J. Wineland, and D. J. Heinzen, Phys. Rev. A \textbf{54}, R4649 (1996).

\bibitem{NOON}
K. Edamatsu, R. Shimizu, and T. Itoh Phys. Rev. Lett. \textbf{89}, 213601 (2002);
A. N. Boto et al., Phys. Rev. Lett. \textbf{85}, 2733 (2000);
M. D'Angelo, M. V. Chekhova, and Y. Shih, Phys. Rev. Lett. \textbf{87}, 013602 (2001);
S. F. Huelga et al., Phys. Rev. Lett. \textbf{79}, 3865 (1997).

\bibitem{NOONth}
S. Boixo, A. Datta, M. J. Davis, A. Shaji, A. B. Tacla, and C. M. Caves, Phys. Rev. A \textbf{80}, 032103 (2009);
Y. M. Hu, M. Feng, and C. Lee, arXiv:1110.1429v2.

\bibitem{NOONexp}
D. Leibfried, M. D. Barrett, T. Schaetz, J. Britton, J. Chiaverini, W. M. Itano, J. D. Jost, C. Langer, and D. J. Wineland, Science \textbf{304}, 1476 (2004);
C. A. Sackett, D. Kielpinski, B. E. King, C. Langer, V. Meyer, C. J. Myatt, M. Rowe, Q. A. Turchette, W. M. Itano, D. J. Wineland, and C. Monroe, Nature (London) \textbf{404}, 256 (2000)

\bibitem{James1998}
D. F. V. James, Appl. Phys. B: Lasers Opt. \textbf{66}, 181 (1998).

\bibitem{Ferdinand2003}
F. Schmidt-Kaler, S. Gulde, M. Riebe, T. Deuschle, A. Kreuter,
G. Lancaster, C. Becher, J. Eschner, H. H\"affner, and R.
Blatt, J. Phys. B \textbf{36}, 623 (2003).

\end{thebibliography}
\end{document}